# Uncovering the Origin of the Emitting States in Bi$^{3+}$-Activated Ca$M$O$_3$ ($M$=Zr, Sn, Ti) Perovskites: Metal-to-Metal Charge Transfer *versus* s-p Transitions

Michele Back,[1,*] Jumpei Ueda,[1] Jian Xu,[1] Kazuki Asami,[1] Lucia Amidani,[2,3] Enrico Trave,[4] Setsuhisa Tanabe[1,*]

[1] Graduate School of Human and Environmental Studies, Kyoto University, Kyoto 606-8501, Japan

[2] Rossendorf Beamline at ESRF – The European Synchrotron, CS40220, 38043 Grenoble Cedex 9, France

[3] Helmholtz-Zentrum Dresden-Rossendorf, Institute of Resource Ecology, P.O. Box 510119, 01314 Dresden, Germany

[4] Department of Molecular Sciences and Nanosystems, Università Ca' Foscari Venezia, Via Torino 155, 30172 Mestre-Venezia, Italy

**ABSTRACT:** After more than a century of studies on the optical properties of Bi$^{3+}$ ion, the assignment of the nature of the emissions and the bands of the absorption spectra remain ambiguous. Here we report an insight into the spectroscopy of Bi$^{3+}$-activated Ca$M$O$_3$ perovskites ($M$=Zr, Sn, Ti), discussing the factors driving the metal-to-metal charge transfer and sp → s$^2$ transitions. With the aim to figure out the whole scenario, a combined experimental and theoretical approach is employed. The comparison between the temperature dependence of the PL emissions with the temperature dependence of the exciton energy of the systems has led to an unprecedent evidence of the charge transfer character of the emitting states in Bi$^{3+}$-activated phosphors. Low temperature VUV spectroscopy together with the design of the vacuum referred binding energy diagram of the luminescent center are exploited to shed light on the origin of the absorption bands. In addition, the X-ray absorption near edge structure, unambiguously confirmed the stabilization of Bi$^{3+}$ in Ca$^{2+}$ site in both CaSnO$_3$ and CaZrO$_3$ perovskites. This breakthrough into the understanding of the excited state origin of Bi$^{3+}$ could pave the way towards the design of a new generation of effective Bi$^{3+}$-activated phosphors.

## INTRODUCTION

Bismuth-based compounds are a well-known family of materials finding applications in many different fields such as water splitting,[1-3] fuel cells,[4] ferroelectrics,[5] cosmetics,[6,7] photovoltaic[8] and optical applications.[9-13] The interest in these compounds lies in the properties induced by the presence of the 6s$^2$ lone electron pair of Bi$^{3+}$ resulting in non-centrosymmetric sites[14] and playing a critical role in the valence band engineering.[2] On the other hand, bismuth-activated luminescent materials have attracted increasing attention owing to their potentialities in a wide range of photonic applications such as new non-lanthanide phosphors, broadband amplifiers and fiber lasers.[15-17] The interesting properties of the Bi-activated luminescent materials arise from the easy involvement in chemical bonds of the *p*-orbitals allowing, in principle, a wide tunability of emissions. Moreover, Bi$^{3+}$ ion is so far one of the most used sensitizers for the enhancement of the luminescent performances of lanthanide activated phosphors[18,19] and persistent luminescent materials.[20,21]

The recent development of highly efficient near-UV laser diode for the design of the next generation phosphor-converted white light-emitting diodes with suitable correlated color temperature and color rendering index, gave a new input to the study of Bi-activated phosphors leading to the development of new materials with high quantum yield (QY) such as La$_2$Zr$_2$O$_7$:Bi$^{3+}$ (QY≈80-90%)[22,23] and La$_4$GeO$_8$:Bi$^{3+}$ (QY≈88%).[24] However, despite extensive efforts dedicated to the understanding of the optical properties of Bi$^{3+}$ ion in solids,[25-29] many fascinating aspects are still debated[30] limiting a rational design of efficient materials. The $^1$S$_0$ ground state (GS) of Bi$^{3+}$ free ion has a 6s$^2$ electronic configuration while the 6s$^1$6p$^1$ configuration gives rise to the triplets $^3$P$_0$, $^3$P$_1$, $^3$P$_2$ and singlet $^1$P$_1$ excited states. Transitions from the $^1$S$_0$ ground state to the excited states $^3$P$_1$, $^3$P$_2$ and $^1$P$_1$ are usually denoted as A, B and C, respectively. If the allowed C-transition is usually located in the VUV region, $^1$S$_0$ → $^3$P$_2$ transitions (B-bands) are spin forbidden, while the $^1$S$_0$ → $^3$P$_1$ transition (A-band) becomes allowed by spin-orbit coupling between $^3$P$_1$ and $^1$P$_1$. The transition from the GS to the lower energy excited state $^3$P$_0$ is strictly forbidden by selection rules but this state is proposed to play a role as a "trap" in the temperature dependence of the $^3$P$_1$ → $^1$S$_0$ transition. Fast and parity allowed 6s$^2$ ↔ 6s$^1$6p$^1$ inter-configurational transitions may lead to efficient excitations and emissions in Bi$^{3+}$-activated materials. However, many different luminescent signals ascribed to Bi-related structures are also frequently present and sometimes difficult to assign.[30,31] The so-called D-state is of particular interest due to the wide emission tunability and it is usually considered to be originated from an impurity trapped exciton state[32] or describe as metal-to-metal charge transfer (MMCT) state.[33]

From the first systematic work of Pohl[34] and the contribution of Seitz,[35] the spectroscopy of the $ns^2$-type ions (e.g. $Sn^{2+}$, $Sb^{3+}$, $Tl^+$, $Pb^{2+}$ and $Bi^{3+}$) were extensively investigated. Ranfagni et al.[36] deeply reviewed the state of the art of the $ns^2$-type ions spectroscopy ($Tl^+$, $Pb^{2+}$, $Bi^{3+}$, $Sn^{2+}$ and $Sb^{3+}$), in particular for $Tl^+$-activated materials, trying to summarize the effects involved in the strong dependence of the absorption and emission due to the host nature and the difficult assignment of the absorption and emission bands. More recently, Blasse, [25,27,33] Boulon,[37,38] Srivastava,[32,39,40] Boutinaud[30,41] and co-workers tried to suggest rule of thumbs and empirical methods for the assignment. However, after more than 100 years of extensive investigation on the spectroscopy of $Bi^{3+}$-activated materials, the assignment of the emission and absorption bands in the photoluminescence (PL) spectra are still ambiguous. Therefore, understanding the important parameters describing the nature of the transitions involved is a key objective for the design of new strategies aimed at overcoming the limitations of the present materials such as undesirable thermal quenching or low QY. Following the pioneering work of Blasse et al.,[42] Kang et al.[17] investigated the tunability of the Bi-related photoemission bands in the $(Y,Sc)(Nb,V)O_4$:$Bi^{3+}$ class of phosphors showing promising properties with a wide emission tunability. However, the interpretation of an $s^2 \rightarrow sp$ transition modulated by the band-to-band of the host proposed by the authors is in contrast with the charge transfer character of the transition reported by Blasse[42] and Boutinaud,[30,41] remaining an unsolved issue. In the same way, the assignment of the broad emission and the relative absorptions in the $C_2$ site of the famous $Y_2O_3$:$Bi^{3+}$ phosphor is also still debated.[30,43] In this context, an insight into the parameters that allow the modulation of the PL output in different bismuth-activated compounds is highly desirable, requiring a detailed investigation to disclose the origin of the transitions involved. In this direction, Ca$M$O$_3$ perovskites are considered an ideal prototype because of the presence of a single $A$ site ($Ca^{2+}$) for $Bi^{3+}$ substitution and the different nature of the PL emission bands demonstrated in this family of compounds.[31,39]

Here, we present a spectroscopic investigation on the luminescent properties of $Bi^{3+}$-activated Ca$M$O$_3$ perovskites ($M$=Zr, Sn, Ti) discussing the origin of the emitting and absorbing states. The whole scenario is supported by the experimental and theoretical analysis employed. An unprecedent correlation between the emitting state and the host behaviour is demonstrated by comparing the temperature dependence of PL emissions with the temperature dependence of the exciton peak of the compounds in a wide temperature range. In addition, combining low temperature VUV-vis spectroscopy together with empirical and theoretical models, the energy level diagram of the luminescent center is designed with respect to the conduction band and valence band energies of the hosts. This approach could be used for a better understanding of the origin of the excited states in Bi-activated luminescent materials and design new effective phosphors.

## RESULTS AND DISCUSSION

**Structural and Optical Properties: Empirical Model of MMCT in Ca$M$O$_3$:$Bi^{3+}$ Perovskites**. XRPD patterns of the samples show the single phase stabilization of the corresponding Ca$M$O$_3$ orthorhombic distorted perovskite structure $Pbmn$ ($M$=Zr, Sn, Ti) for all the samples (Figure 1a). No other peaks are detected, confirming the goodness of the synthesis and the successful introduction of the doping cations into the perovskite structure. Orthorhombic distorted perovskites of $A^{2+}M^{4+}O_3$ formula are often referred to be GdFeO$_3$ type structure. The rotation of the $M$O$_6$ octahedra are about the diad axes and it corresponds to a tilt group $a^+b^-b^-$ (Ref. 44; Glazer tilt notation). Figure 1b shows the crystal structure of a generic orthorhombic distorted Ca$M$O$_3$ perovskite.

Belonging to the same distorted orthorhombic crystal structure family with a single $Ca^{2+}$ site for the substitution of $Bi^{3+}$, this class of compounds is considered an ideal prototype for the investigation of the host effect on the excited states nature of $Bi^{3+}$ and their optical properties. Figure 2a and b show the comparison of the reflectance spectra and the typical luminescent features of the Ca$M$O$_3$:$Bi^{3+}$ systems ($M$=Zr, Sn, Ti), respectively. The PL emissions display a clear red-shift trend of the emission band moving from CaZrO$_3$ to CaSnO$_3$ and CaTiO$_3$, in agreement with the optical behaviour reported in the literature.[31,39,45] CaSnO$_3$:$Bi^{3+}$ photoluminescence excitation (PLE) spectrum is composed of two bands, with maximum peaks at 258 nm and 307 nm. The peak at higher energy (4.79 eV) can be assigned to the host exciton absorption.[46] The strong absorption by host exciton indicates efficient energy transfer process from the host to the activator. The PLE spectrum of CaTiO$_3$:$Bi^{3+}$ consists of two absorption bands, one placed at 328 nm, assigned to the excitonic peak of the host, and another broad band at lower energy related to Bi. By considering (i) the large Stokes shift $\Delta E_{Stokes}$ (1.33 eV and 1.46 eV in CaSnO$_3$ and CaTiO$_3$, respectively) between the absorption and the emission peaks, (ii) the low emission energy with respect to the typical UV emissions of $Bi^{3+}$-activated oxide materials and (iii) the large full width at half maximum (FWHM) of 0.66 eV and 0.53 eV for CaSnO$_3$:$Bi^{3+}$ and CaTiO$_3$:$Bi^{3+}$, respectively, we assign the emission peaks of both the CaSnO$_3$:$Bi^{3+}$ and CaTiO$_3$:$Bi^{3+}$ to the D-state, in agreement with previous

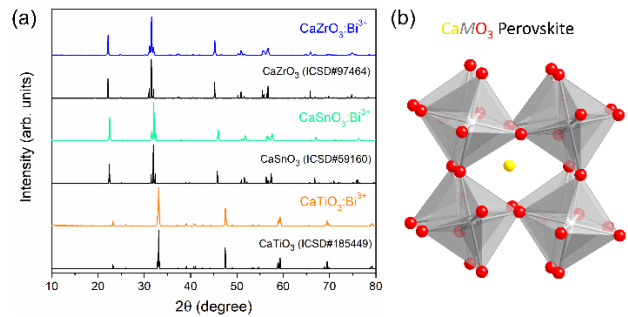

**Figure 1.** (a) XRPD patterns of the Ca$M$O$_3$:$Bi^{3+}$ ($M$=Zr,Sn,Ti) samples and (b) crystalline structure of a generic Ca$M$O$_3$ orthorhombic distorted perovskite characterized by a tilt group $a^+b^-b^-$.

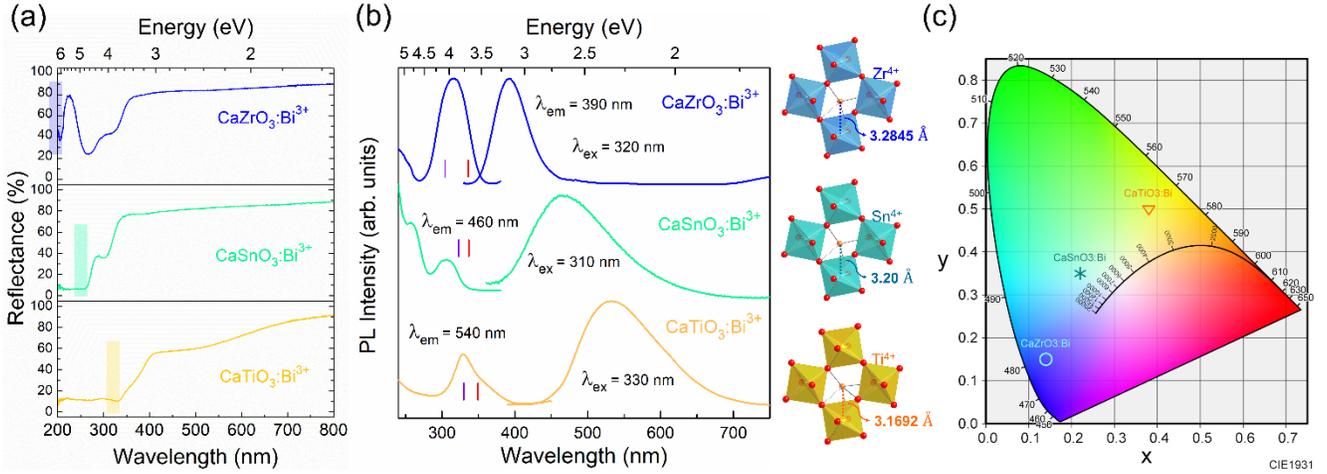

**Figure 2.** Reflectance (a) and PL and PLE (b) spectra for the Ca$MO_3$:Bi$^{3+}$ systems ($M$=Zr, Sn, Ti) at room temperature. The vertical violet and red lines represent the MMCT and A transitions estimated from the empirical models using Eq. 1 and Eq. 3, respectively. The typical structure of the three perovskites under consideration is depicted by underlying the shortest distance between the Ca$^{2+}$ site and the nearest neighbor $M^{4+}$ ion. (c) Chromaticity diagram showing the tunability of the color coordinates of the investigated materials.

observations.[31,39]

Parameters such as the Stokes shift $\Delta E_{Stokes}$, the emission energy and the FWHM of the PL can be considered to be a good starting point for the assignment of the emission peak origin. In terms of all these parameters it is clear the difference in PL spectral feature between the PL emission in CaZrO$_3$, assigned to the Bi$^{3+}$ $^3P_{0,1} \rightarrow$ $^1S_0$ transition, with respect to that one in CaSnO$_3$ and CaTiO$_3$, assigned to the MMCT state.

The change in the $M$ site of perovskite moving from Zr$^{4+}$ to Sn$^{4+}$ and finally to Ti$^{4+}$, gives rise to a big change in the emission color output of the samples as evidenced by the chromaticity diagram (CIE1931) in Figure 2c. Table 1 summarizes the CIE color coordinates $x$ and $y$ and the maximum wavelength $\lambda_{max}$ characterizing the emission bands of the Ca$MO_3$:Bi$^{3+}$ systems.

If the emission peaks can be assigned with a certain confidence by means of this rule of thumb, the unequivocal identification of the nature of the excitation peaks is much more complicated, requiring a deeper investigation. As a first attempt, by judging from the difference in the PLE spectral shape of the lowest lying transition state, the s-p transition nature in CaZrO$_3$ and the charge transfer nature of the broad band centred at about 350 nm in CaTiO$_3$ can be inferred. Concerning Bi$^{3+}$ in CaSnO$_3$, the shape suggests the 6s$^1$6p$^1$ character of the excited state, however, this assignment is ambiguous.[39,41,47]

For a better understanding of the complicate nature of the absorption bands of Bi$^{3+}$-based compounds, semi-empirical models for the theoretical calculation of the A and C bands and the MMCT excited states were proposed in the last decade.[30,48] By a detailed analysis of experimental data reported in the literature, Boutinaud et al.[30] proposed a semi-empirical model to predict the energy of MMCT transitions in oxide compound containing Bi$^{3+}$. For a $M^{n+}$ with coordination numbers larger than 4 (as in the present case), the following empirical equation was proposed:

$$E_{MMCT,th}(\text{cm}^{-1}) = 55000 - 45500 \frac{\chi_{CN'}(M^{n+})}{d_{corr}}. \quad (1)$$

$\chi_{CN'}(M^{n+})$ is the electronegativity (considering the new scale reported by Li and Xue[49]) for the host cations $M^{n+}$ with coordination number CN' and $d_{corr}$ is the shortest distances between the $M^{n+}$ site and the cation site available for Bi$^{3+}$ (Ca$^{2+}$ in this work), corrected to the anion relaxation effect by means of the following formula:[50]

$$d_{corr} = d_{host} + \frac{1}{2}[r(\text{Bi}^{3+}) - r(\text{host})] \quad (2)$$

where $r(\text{Bi}^{3+})$ is the Bi$^{3+}$ ionic radius and $r(\text{host})$ is the ionic radius of the host cation to be substituted for Bi$^{3+}$ ion. With this model, the absorption into the MMCT state (D-state) is predicted to occur at about 4.08, 3.84, and 3.76 eV in the Ca$MO_3$ perovskites moving from $M$=Zr, to Sn and finally to Ti, respectively (Table 2).

Wang et al.[48] demonstrated the possibility to empirically estimate the energy of the A and C transitions ($E_{A(Bi)}$ and $E_{C(Bi)}$), based on the following equations:

$$E_{A(Bi)}(X, \text{cm}^{-1}) = 23970 + 50051\, e^{-\frac{h_e}{0.551}} \quad (3)$$

$$E_{C(Bi)}(X, \text{cm}^{-1}) = 26100 + 88100\, e^{-\frac{h_e}{0.644}} \quad (4)$$

**Table 1** Color coordinates ($x$ and $y$), maximum emission wavelength ($\lambda_{max}$) at room temperature and quenching temperature $T_{50}$ for the Ca$MO_3$:Bi$^{3+}$ perovskite systems ($M$=Zr, Sn and Ti).

| Perovskite | $x$ | $y$ | $\lambda_{max}$ (nm) | $T_{50}$ (K) |
|---|---|---|---|---|
| CaZrO$_3$:Bi$^{3+}$ | 0.14 | 0.15 | 390 | 380 |
| CaSnO$_3$:Bi$^{3+}$ | 0.22 | 0.35 | 460 | 225 |
| CaTiO$_3$:Bi$^{3+}$ | 0.38 | 0.50 | 535 | 210 |

**Table 2** Theoretical energy values of the MMCT ($E_{MMCT,th}$), A ($E_{A(Bi)}$) and C ($E_{C(Bi)}$) states from the semi-empirical models for the Ca$M$O$_3$ perovskites doped with bismuth.

| Perovskite | $\chi_{CN'}(M^{n+})$[a] | $d_{corr}$[b] (Å) | $E_{MMCT,th}$[c] (eV) | $h_e$[d] | $E_{A(Bi)}$[e] (eV) | $E_{C(Bi)}$[f] (eV) |
|---|---|---|---|---|---|---|
| CaZrO$_3$:Bi$^{3+}$ | 1.610 | 3.3095 | 4.08 | 1.19 | 3.69 | 4.96 |
| CaSnO$_3$:Bi$^{3+}$ | 1.706 | 3.2250 | 3.84 | 1.20 | 3.68 | 4.93 |
| CaTiO$_3$:Bi$^{3+}$ | 1.730 | 3.1942 | 3.76 | 1.31 | 3.55 | 4.67 |

[a] electronegativity values from ref. 49; [b] calculated from Eq. 2 considering the ICSD#97464, 9015771 and 185449 for CaZrO$_3$, CaSnO$_3$ and CaTiO$_3$, respectively; [c] calculated from Eq. 1; [d] calculated from Eq. 5; [e] calculated from Eq. 3; [f] calculated from Eq. 4.

which depend on the environmental factor $h_e(X)$ defined as:

$$h_e(X) = \sqrt{\sum_1^{N_X} f_c(X-L)\, \alpha(X-L)\, Q_L^2} \quad (5)$$

where $f_c(X-L)$ and $\alpha(X-L)$ are the fractional covalency and volume polarization of each bond $X-L$ in binary units of the host lattice, respectively, and $Q_L$ is the effective charge of the ligand (here oxygen) in each unit. A detailed explanation of the parameters can be found in refs. 51 and 52.

Table 2 summarizes the values estimated for the three perovskites and the red and violet vertical lines in Figure 2b represent the values estimated from the theories for the A and MMCT states respectively. Looking at the values, the empirical models predict a situation in which the low-lying excited state is always the $^3P_{0,1}$ state, in contradiction with the MMCT nature of the broad band in CaTiO$_3$.[31] In addition, if we only consider the difference of the values of the calculated MMCT and the A states in comparison to the energy of the absorption peak in the PLE spectrum of Bi$^{3+}$ in CaSnO$_3$ (4.04 eV) and in CaZrO$_3$ (3.92 eV), it is difficult to unequivocally assign the origin of the transitions, then the MMCT state could be possible in both the cases.

**Temperature dependence of PL.** From the fundamental and the application point of view, important information can be derived from the effect of the temperature on the PL emission behaviour. For a comprehensive investigation, $ns^2$ ions required a spectroscopic analysis in a wide temperature range, starting from low temperatures. The results of the thermal effect are summarized by the PL spectra shown in Figure 3.

The temperature dependence of the PL emission spectra in CaZrO$_3$:Bi$^{3+}$ system was investigated in the 20-600 K range (Figure 3a). The analysis was carried out on the energy scale spectra, converted by means of the Jacobian transformation. The plot of the integrated area of the emission band as a function of temperature, reported in Figure 3b, cannot be reproduced by means of the simple crossover quenching process described by Struck and Fonger model.[53] The two different evidenced processes can be described by the following equation:

$$I(T) = \frac{I(0)}{1 + C_1 exp(-\Delta E_1/k_B T) + C_2 exp(-\Delta E_2/k_B T)}. \quad (6)$$

The solid line in Figure 3b represents the best fit of the data (circles). The fit allowed to estimate the activation energies $\Delta E_1$ (165±17 cm$^{-1}$) and $\Delta E_2$ (2820±110 cm$^{-1}$) of the quenching processes.

The temperature dependence of the PL emission in CaSnO$_3$:Bi$^{3+}$ was analyzed in the 20-550 K temperature range showing the same unusual behaviour. Figure 3d shows the PL spectra as a function of temperature with an evident decrease of the emission at 460 nm from the MMCT state at increasing temperature. As for the case of CaZrO$_3$:Bi$^{3+}$, the temperature dependence of the integrated area cannot be fit by a single-quenching process. This behaviour indicates that more processes contribute to the luminescence quenching and two different regions can be distinguished: (i) up to about 250 K, a first quenching process is shown, and (ii) from about 300 K, a further decrease of the MMCT emission at increasing temperature can be evidenced. The best fit using Eq. 6 allows to estimate the activation energies $\Delta E_1$ (200±10 cm$^{-1}$) and $\Delta E_2$ (3420±330 cm$^{-1}$) of the quenching processes evidenced for the MMCT emission. It is interesting to note how this unusual temperature behaviour shown by the PL intensity of Bi$^{3+}$ in both the CaZrO$_3$ and CaSnO$_3$ perovskites is similar to the one reported for others Bi$^{3+}$-activated compounds such as Bi$_4$Ge$_3$O$_{12}$,[54] LaPO$_4$:Bi$^{3+}$ [55] and LaGaO$_3$:Bi$^{3+}$.[56] The nature of this unconventional behaviour was explained by involving three possible processes: (1) the $^3P_0$ state lies below the $^3P_1$ working as a "trap"[25] and inducing a thermal behaviour that is explained by a three-energy level scheme,[37,38] (2) a pseudo Jahn-Teller effect resulting in the A$_T$ and A$_X$ emission bands[36,57] or (3) D-band effect.[36] Blasse et al.[25,57] analyzed the strong dependence of the host lattice on the luminescence properties of Bi$^{3+}$ ions for oxides and halides demonstrating that a small Stokes-shift (and vibrational structure) is only observed for 6-fold coordinated Bi$^{3+}$, while in 8-fold or higher coordination it has been proposed that Bi$^{3+}$ in the ground state is characterized by off-center preferring an asymmetric coordination. The result is a large Stokes-shift originated from a large relaxation. This agrees with the fact that no vibrational structure was detected at low temperature.

At the present stage, this double quenching process is still not perfectly clear and a detailed investigation is necessary to unveil its real origin. Nevertheless, it is interesting to compare the effect of the temperature on the emission of Bi$^{3+}$ in the different perovskites and the quenching temperature $T_{50}$ (temperature at which the intensity reaches 50% of the intensity at low temperature), can be considered as an effective parameter. From the temperature dependence of the integrated PL emission trend of the CaZrO$_3$:Bi$^{3+}$ and CaSnO$_3$:Bi$^{3+}$ compounds, quenching temperature $T_{50}$ of 380 K and 225 K are estimated, respectively. Srivastava et al.[58] estimated a quenching activation energy of 645 cm$^{-1}$ for Bi$^{3+}$ in CaZrO$_3$, however, the analysis reported is limited into the 10-300 K range with a decrease of the low temperature intensity of only the 25% at 300 K ($T_{75}$), in agreement with Figure 3b.

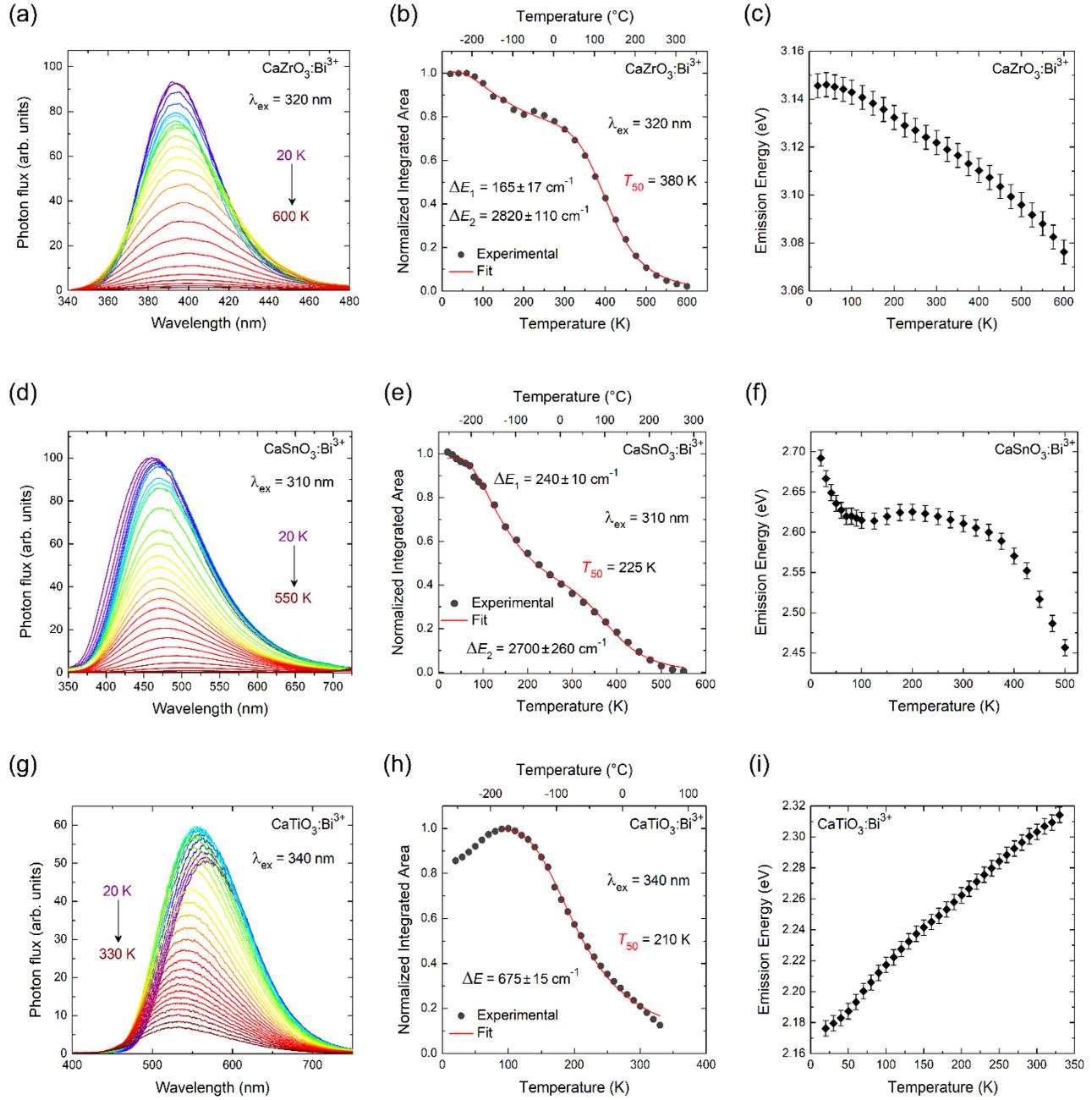

**Figure 3.** Temperature dependence of PL spectra (a,d,g), integrated area (b,e,h) and emission band energy (c,f,i) of the Ca$M$O$_3$:Bi$^{3+}$ samples ($M$=Zr, Sn, Ti). The spectra were collected in the 20-600 K range exciting at 320 nm, 20-550 K range exciting at 310 nm and 20-330 K range exciting at 340 nm for CaZrO$_3$:Bi$^{3+}$, CaSnO$_3$:Bi$^{3+}$ and CaTiO$_3$:Bi$^{3+}$, respectively. The fit using Eq. 6 is reported as red curves for CaZrO$_3$:Bi$^{3+}$ (b) and CaSnO$_3$:Bi$^{3+}$ (e) while the single barrier quenching model fit using Eq. 7 is employed for CaTiO$_3$:Bi$^{3+}$ (h).

To estimate the $T_{50}$, the temperature dependence of the CaTiO$_3$:Bi$^{3+}$ system was investigated in the 20-330 K range (Figure 3g). The activation energy $\Delta E$ of the quenching process in CaTiO$_3$:Bi$^{3+}$ can be calculated by means of the single-barrier model:[53]

$$I(T) = \frac{I(0)}{1 + (\Gamma_0/\Gamma_v)e^{-\Delta E/k_B T}} \quad (7)$$

where $I$ represents the PL intensity, $\Gamma_v$ the radiative rate, $\Gamma_0$ the attempt rate of the nonradiative process, $k_B$ the Boltzmann constant and $T$ the temperature. The activation energy was estimated to be $\Delta E$ = 2442±112 cm$^{-1}$ and the $T_{50}$= 210 K. Boutinaud and Cavalli[31] reported a severe quenching process for the emission coming from the CaTiO$_3$:Bi$^{3+}$ system, estimating a small activation energy $\Delta E$ of 300 cm$^{-1}$ and a $T_{50}$=140 K can be extrapolated. This discrepancy is due to the high concentration of Bi (5%) used by the authors. As a result, the following general trend is found: $T_{50}(\text{CaZrO}_3:\text{Bi}^{3+}) > T_{50}(\text{CaSnO}_3:\text{Bi}^{3+}) > T_{50}(\text{CaTiO}_3:\text{Bi}^{3+})$.

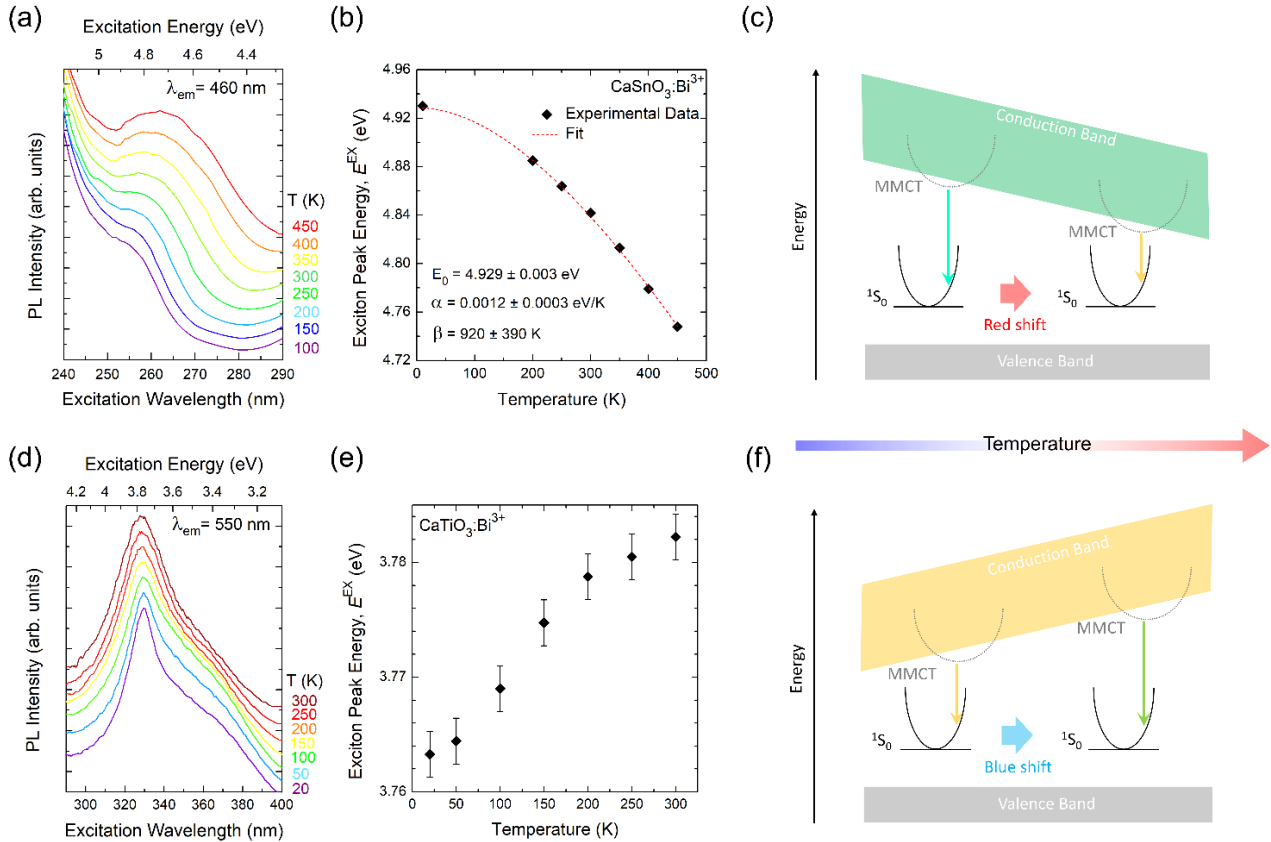

**Figure 4.** Temperature dependence of PLE spectra of the exciton peak (a,d), trend of the exciton peak energy $E^{EX}$ as a function of temperature (b,e) and sketch of the temperature effect on the red and blue shift of the CaSnO$_3$ and CaTiO$_3$ host excitons on the MMCT red and blue shift, respectively (c,f). The red dashed curve in (b) represents the fit using the Varshni equation (Eq. 8).

Considering Table 1 and 2, the red-shift emission trend observed in the PL spectra seems to be related to the electronegativity $\chi_{CN'}(M^{n+})$ of the $M^{4+}$ cation and the $d_{corr}$ distance, with an increase of the emission wavelength (decrease of the energy) at the increasing $\chi_{CN'}(M^{n+})$ and the consequent decreasing distances between Bi$^{3+}$ site and $M^{4+}$ in the structures. However, by decreasing the Bi$^{3+}$-$M^{4+}$ nearest neighbour distance, the $T_{50}$ becomes progressively lower.

In addition to the intensity, critical information can be obtained from the analysis of the temperature dependence of the emission bands shape and position. Figure S1 shows the temperature evolution of the normalized PL spectra for the three Ca$M$O$_3$:Bi$^{3+}$ systems and Figure 3c, f and i summarize the temperature dependence of the emitting bands energy. Bi$^{3+}$-doped compounds characterized by the $^3P_1 \rightarrow {}^1S_0$ transition, usually show a typical blue shift of the emitting band by increasing the temperature.[58-61] In contrast with this trend, the energy of the emitting peak of Bi$^{3+}$ in CaZrO$_3$ evidences a clear red shift at temperature higher than 100 K. However, the temperature dependence of the normalized PL spectra (Figure S1a) shows the relative increase of a second broad band at lower energies with the increase of the temperature, explaining the red shift. This second broad band could be assigned to the emission from the MMCT. With this assumption, the two-quenching process in Figure 3b can be explained by the thermal population of the MMCT state at the expenses of the $^3P_1$ emission and, subsequently, the MMCT quenching.

**Temperature dependence of Exciton Peaks.** As previously discussed, the emission bands of CaSnO$_3$:Bi$^{3+}$ and CaTiO$_3$:Bi$^{3+}$ are assigned to the MMCT state. Because of the charge transfer nature of the emitting state, that is strongly related to the conduction band bottom of the host, a temperature dependence linked to the bandgap behavior is expected. Density functional theory calculations of the density of states (DOS) of CaSnO$_3$ and CaTiO$_3$ have demonstrated the bottom of the conduction band (CB) is mainly formed by the 6s, 6p-Sn orbitals[62] and the 3d-Ti orbitals,[63] respectively. In this context, the temperature dependence of the bandgap is usually characterized by a red shift by increasing the temperature[64-66] and a consequent red shift is expected for the MMCT emission band rising the temperature. However, if for CaSnO$_3$:Bi$^{3+}$, a general red shift is measured (Figure 3f), CaTiO$_3$:Bi$^{3+}$ shows the opposite trend with a clear blue shift with the temperature (Figure 3i). In the normalized spectra of CaTiO$_3$:Bi$^{3+}$ system (Figure S1b), no secondary peak can be disclosed. Therefore, the thermal response of Bi$^{3+}$-related luminescence in CaTiO$_3$ is apparently not following the expected trend.

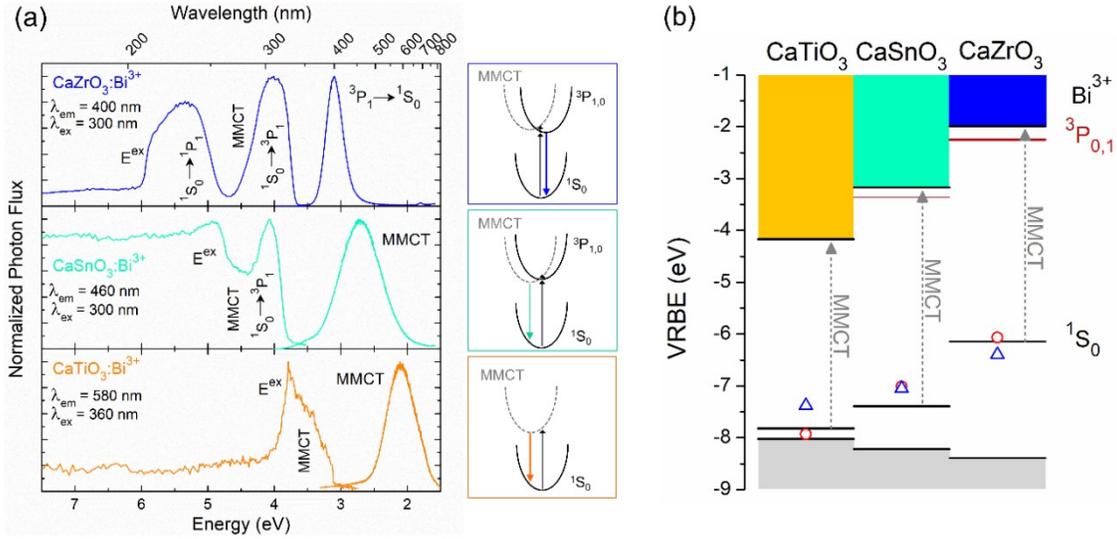

**Figure 5.** (a) Synchrotron radiation VUV-vis PL and PLE spectra for Ca$M$O$_3$:Bi$^{3+}$ systems ($M$=Zr, Sn, Ti) collected at 9 K and a schematic configurational coordinate diagram for the Bi$^{3+}$ states and MMCT. (b) VRBE diagram constructed by experimental and theoretical data. The red circles and the blue triangles represent the $^1S_0$ ground state energy estimated from the theoretical values calculated for the MMCT-state absorption using Eq. 1 and from the values extrapolated from the $^1S_0$ vs $U(6,A)$ trend reported in ref. 47, respectively.

In order to shed light on this unexpected behavior, the temperature dependence of the exciton peak (and consequently of the bandgap) of CaSnO$_3$ and CaTiO$_3$ was investigated (Figure 4). The temperature dependence of the PLE spectra in Figure 4a and d and the trend of the resulting exciton peak energy as a function of temperature in Figures 4b and e clearly show an opposite behavior. As for most semiconductors, the temperature dependence of the energy bandgap of CaSnO$_3$ (or equivalently the exciton peak) can be modeled by the Varshni equation:[64]

$$E^{EX}(T) = E^{EX}(0) - \frac{\alpha T^2}{T+\beta} \quad (8)$$

where $E^{EX}(0)$ is the exciton peak energy at $T$=0 and $\alpha$ and $\beta$ are constants, with $\beta$ proportional to the Debye temperature.[64,67] The dashed line in Figure 4b represents the best fit, resulting in $E^{EX}(0)$ = 4.929±0.003 eV, $\alpha$ = 0.0012±0.0003 eV·K$^{-1}$ and $\beta$ = 920±390 K. The bandgap energy $E_g(0)$ extrapolated at $T$=0 can be calculated considering the relation $E_g = E^{EX} + 0.008 \times (E^{EX})^2$,[68] giving a value of 5.12 eV. This trend agrees with the red shift behavior of the emission band energy in CaSnO$_3$:Bi$^{3+}$. In the same way, the blue shift of the PL emission in CaTiO$_3$:Bi$^{3+}$ system is in agreement with the blue shift of the exciton peak (Figure 4e). The temperature dependence of the $E_g$ arises from the thermal lattice expansion and the electron-phonon interactions.[65,66] In polar and partially ionic crystals, an additional contribution due to the Fröhlich interaction plays an important role and in case of titanates such as CaTiO$_3$, the temperature dependence of the $E_g$ was demonstrated to be governed by this interaction.[69]

The comparison of the temperature dependence of the emission band energy with the trend of the exciton peak energy suggests a strong connection between the emitting state and the host band structure, supporting the assignment of the charge transfer character of the emitting state. The sketches in Figures 4c and f represent the effect of the bandgap shift with the temperature to the MMCT emission energy shift showing an unprecedent correlation between the MMCT state and the host.

**Low Temperature VUV Spectroscopy.** In the attempt to provide a comprehensive overview, low temperature synchrotron radiation VUV-Vis spectra were collected. Figure 5a shows the energy scale PL and PLE spectra at 9 K of the samples from 1.5 to 7.5 eV. The PL spectra show the expected red-shift with a clear increase of the FWHM for the emission band of CaSnO$_3$ and CaTiO$_3$ (0.72 and 0.52 eV) with respect to the emission in CaZrO$_3$ (0.28 eV), in agreement with the MMCT character of the transition with respect to the $^3P_{1,0} \rightarrow {}^1S_0$. In addition, the impossibility to observe the vibrational structure in the Bi$^{3+}$ spectra of all the perovskites (8-fold or higher coordination) agrees with the observation of the vibrational structure only for 6-fold coordinated Bi$^{3+}$.[57] The PLE spectra of the CaTiO$_3$:Bi$^{3+}$ system is characterized by the host exciton peak at 3.79 eV and the MMCT absorption band centered at 3.65 eV with a Stokes shift $\Delta E_{Stokes}$=1.55 eV. By taking advantage of the defect emission of CaTiO$_3$, the host exciton peak can be well separated from the MMCT excitation peak (Figure S2a). CaZrO$_3$:Bi$^{3+}$ compound exhibits a more structured excitation spectrum in which four bands can be distinguished. The deconvolution allows to estimate the maximum of the bands at 5.70 eV, 5.23 eV, 4.15 eV and 3.89 eV, assigned to the host band, the $^1S_0 \rightarrow {}^1P_1$ (C transition), MMCT and $^1S_0 \rightarrow {}^3P_1$ (A transition), respectively (deconvolution in Figure S2b). Remarkably, by considering an energy shift of about 0.2 eV due to the low temperature measurements, the energies estimated from the semi-empirical models in Table 2, well match with the values estimated experimentally. The discrepancy is within the accuracy of ±0.37 eV for the prediction of the MMCT energies.[30] Moreover, a Stokes shift of 0.79 eV is estimated for the s-p transition. In the case of CaSnO$_3$:Bi$^{3+}$, two peaks are clearly detected in the PLE. However, with only these two peaks, it is impossible to fit the spectra and a third band is necessary (deconvolution in Figure S2c). The highest energy peak due to the host exciton lies at 5.05 eV, while, based on the bandwidth of the fitted peaks, the bands centered at 4.22 eV and 4.03 eV are reasonably assigned to the MMCT and $^1S_0 \rightarrow {}^3P_1$ absorptions, respectively. From the deconvolution of the spectra, the s-p transition for CaSnO$_3$ is characterized by an emission band centered at 3.57 eV and, consequently, a Stokes shift of 0.46 eV. As for CaZrO$_3$, the s-p transition is characterized by a small Stokes shift; otherwise the MMCT state shows a Stokes shift $\Delta E_{Stokes}$=1.52 eV, very close to that one of CaTiO$_3$.

**Table 3** Experimental and calculated energy values of $U(6,A)$, $^3P_1$ and $^1S_0$ states of $Bi^{3+}$, MMCT state ($E_{MMCT,exp}$), $\Delta E_{Stokes}$ and VB and CB for the Ca$M$O$_3$ perovskites doped with bismuth.

| Perovskite | $U(6,A)^a$ | $E_{^3P_1}{}^b$ | $E_{^1S_0}{}^b$ | $E_{A(Bi)}{}^b$ | $E_{MMCT,exp}$ | $\Delta E_{Stokes}$ | $E_V{}^c$ | $E_C{}^c$ |
|---|---|---|---|---|---|---|---|---|
| CaZrO$_3$:Bi$^{3+}$ | 6.55 | -2.70 | -6.40 | 3.70 | 4.15 | 0.79 | -8.39 | -1.99 |
| CaSnO$_3$:Bi$^{3+}$ | 6.75 | -2.95 | -7.05 | 4.10 | 4.22 | 0.46; 1.52 | -8.22 | -3.17 |
| CaTiO$_3$:Bi$^{3+}$ | 6.85 | -3.08 | -7.38 | 4.00 | 3.65 | 1.55 | -8.02 | -4.07 |

$^a$ value from refs. 70 and 71; $^b$ extrapolated from the trend reported in ref. 47; $^c$ VB and CB energies of CaZrO$_3$ from ref. 70 and of CaSnO$_3$ and CaTiO$_3$ from ref. 71.

**VRBE Diagram.** To provide a detailed explanation about the effect of the different $M^{4+}$ cation choice on the optical properties of Bi$^{3+}$ in Ca$M$O$_3$ perovskite family and to figure out the whole scenario, the Dorenbos and Rogers approach based on lanthanide spectroscopy[72] is considered. The energies of the top of the valence band (VB) and the bottom of the CB used in the construction of the vacuum referred binding energy (VRBE) diagram of Figure 5b are summarized in Table 3. The energy of the MMCT state represents the gap between the Bi$^{3+}$ ground state $^1S_0$ and the bottom of the conduction band, allowing the location of the Bi$^{3+}$ ground state in the VRBE diagram ($^1S_0$ ground state from the experimental MMCT-state absorption, theoretically estimated from Eq. 1 and extrapolated from ref. 47 are described as a line, a red circle and a blue triangle, respectively). Moreover, where visible by experiments, the first excited state $^3P_1$ can also be recognized. The VRBE diagrams of the Bi$^{3+}$ with respect to the MMCT-state for the three perovskites are depicted in Figure 5b.

As previously discussed, MMCT states are characterized by large Stokes shift and FWHM, indicating such kind of character for CaSnO$_3$ and CaTiO$_3$, as shown in Figure 5a, while CaZrO$_3$ emission can be ascribed to the first excited state $^3P_1$ of Bi$^{3+}$. By considering the trend of the $^1S_0$ and $^3P_1$ energies with respect to the $U(6,A)$ reported by Awater and Dorenbos,[47] the energies of these two levels were extrapolated and compared with the values estimated from the construction of the VRBE diagram from the experimental data (Figure 5b). From this analysis, in regard to the band structure of the Ca$M$O$_3$:Bi$^{3+}$ systems ($M$= Zr, Sn, Ti), it can be inferred that the electronegativity $\chi_{CN'}(M^{n+})$ and $d_{corr}$ values of CaZrO$_3$ system (defined in Eq. 1) are the highest and lowest threshold values, respectively, for the stabilization of the MMCT state with an emission energy lower than the s-p emission from the $^3P_1$ excited state.

However, as suggested by the analysis of Boutinaud and Cavalli,[30,31] the ratio between the electronegativity and the Bi$^{3+}$-$M^{4+}$ distance, $\chi_{CN'}(M^{n+})/d_{corr}$ seems to be a more reliable parameter to describe the trends in Bi-activated phosphors and ratios of 0.49, 0.53 and 0.54 are estimated in CaZrO$_3$, CaSnO$_3$ and CaTiO$_3$, respectively. Setlur and Srivastava[40] reported a Bi$^{3+}$ emission due to the $^3P_{0,1} \rightarrow {}^1S_0$ transition in CaHfO$_3$ perovskite, with a very similar energy and Stokes shift to that one in CaZrO$_3$. Hf$^{4+}$ is characterized by an electronegativity of 1.706 [49] and, considering the crystal structure analysis reported by Feteira et al.,[73] a $d_{corr}$=3.2930 Å is calculated; consequently, $\chi_{CN'}(M^{n+})/d_{corr}$=0.52. Based on these values, we can argue that a $\chi_{CN'}(M^{n+})/d_{corr}$ threshold lies between 0.52 and 0.53: for values ≤0.52, the $^3P_1 \rightarrow {}^1S_0$ transition dominates the PL spectrum and for ratio ≥0.53, the MMCT is the emitting state. From the structural point of view, the distorted orthorhombic perovskite structure is characterized by a short distance between the $M^{4+}$ ion site and the Bi$^{3+}$ substituting site of Ca$^{2+}$ ion, promoting the process.

By considering the interesting results obtained for the Ca$M$O$_3$ perovskites, we summarized the energy of the emission bands of Bi$^{3+}$-activated phosphors as a function of the $\chi_{CN'}(M^{n+})/d_{corr}$ parameter for a wide set of oxide compounds (Table S1). Figure 6 shows the goodness of $\chi_{CN'}(M^{n+})/d_{corr}$ parameter to predict the origin of the emitting state, confirming the threshold discussed in perovskites, an important starting point for the development of new effective Bi-doped materials.

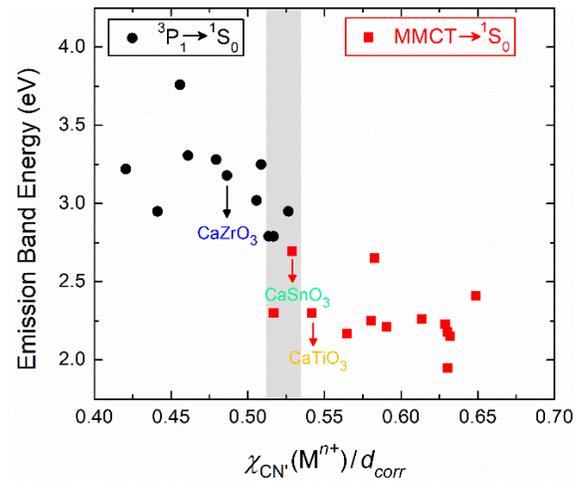

**Figure 6.** Energy of Bi-related emission bands as a function of $\chi_{CN'}(M^{n+})/d_{corr}$ for a set of oxide compounds (values in Table S1). Black dot and red squares represent the emission energy from the $^3P_1$ and MMCT states, respectively.

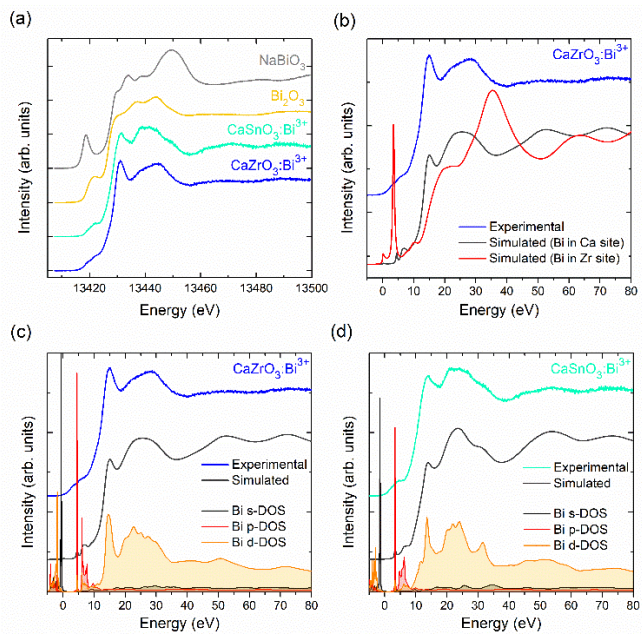

**Figure 7.** (a) Comparison of the Bi $L_3$-edge HERFD-XANES spectra of $CaZrO_3:Bi^{3+}$ and $CaSnO_3:Bi^{3+}$ systems with the $Bi_2O_3$ and $NaBiO_3$ references. (b) Comparison between the experimental and the simulation results in $CaZrO_3:Bi^{3+}$ sample by substituting Bi in Ca or Zr site. Experimental Bi $L_3$-edge HERFD-XANES together with the results of FDMNES calculations and the partial DOS extracted for Bi in $CaZrO_3$ (c) and $CaSnO_3$ (d).

**XANES Spectroscopy.** To confirm the valence state and probe the site of substitution of $Bi^{3+}$ ion, X-ray absorption near edge structure (XANES) analysis was employed. High energy resolution fluorescence detection (HERFD) XANES is a powerful technique to probe the local environment within several angstroms around the investigated atom, particularly suitable in many fields of materials science. Amidani et al.[74] recently applied this technique to investigate the degradation of the blue luminescence of $Eu^{2+}$ doped $BaMgAl_{10}O_{17}$ due to the irradiation. Van der Linden et al.[75] exploited HERFD-XANES spectroscopy to successfully investigate the single Au atom into Ag clusters. Unlike in the field of Bi-based glass, where XANES at the Bi $L_3$ is used to investigate the local environment of Bi ions, this technique is rarely considered in the field of Bi-activated crystalline phosphors. Van Zon et al.[76] revealed the off-centered coordination of $Bi^{3+}$ in $LaPO_4$, due to the presence of the lone pairs, by means of extended X-ray absorption fine structure, then demonstrating the potential of such family of techniques. However, to the best of our knowledge, there is no literature about the use of HERFD-XANES spectroscopy to probe $Bi^{3+}$ ions added as luminescent centers into crystalline phosphors.

Recently, Mistonov et al.[77] demonstrated the clear advantage of measuring XANES in the high energy resolution mode compared to the traditional total fluorescence mode to investigate bismuth-based compounds such as metallic Bi, $Bi_2O_3$, $BiPO_4$, $Bi_4(GeO_4)_3$ and $NaBiO_3$. In addition, the profiles of the Bi $L_3$-edge spectra were demonstrated to significantly differ among different Bi-based compounds in terms of energy shift and spectral shapes,[77] being sensitive not only to the oxidation state but also to both the ligand environment and the crystal structure. By considering the different nature of the emitting transitions of $Bi^{3+}$ in $CaZrO_3$ and $CaSnO_3$, XANES analysis was carried out to investigate the local Bi environment. Bi $L_3$-edge HERFD-XANES spectra for both $CaSnO_3:Bi^{3+}$ and $CaZrO_3:Bi^{3+}$ samples were recorded and compared to first-principles calculations performed with the FDMNES code (Figure 7c and d). The spectra of the two samples are very similar and show a pre-edge structure, the absorption edge, a sharp peak followed by a broader feature and finally a relatively flat post-edge. The main differences appear in the relative intensities of the two features after the absorption edge. Comparison with the references shows that the pre-edge in $CaZrO_3$ and $CaSnO_3$ are at the same energy of the pre-edge of $Bi_2O_3$, but while the latter is a single intense feature the pre-edge of the samples is made by two components. The edge position is closer to that of $NaBiO_3$ rather than to $Bi_2O_3$. Relative shifts of the absorption edge are often correlated to a change in the valence state of the absorber, however chemical and structural effect are deeply interconnected in XANES and not always this correlation can be made. In the case of Bi, the $L_3$ edge is very sensitive to the local structure while in the $L_1$ edge a systematic shift in accord with the valence state change is found.[77] The features after the post edge differ from both references and are relate to the local structure of Bi dopants.

To better understand the origin of feature observed, we performed first principle calculations with the FDMNES code. From the viewpoint of ionic radius, $Bi^{3+}$ is expected to substitute for $Ca^{2+}$. However, to exclude the possibility of substitution in the $M^{4+}$ site, we firstly compared the effect of the substitution of Bi in Ca and Zr sites ($Zr^{4+}$ is the largest among the $M^{4+}$ cation considered here). The simulations shown in Figure 7b clearly confirmed the stabilization of $Bi^{3+}$ in $Ca^{2+}$ site. Therefore, for the calculations, we substituted one Ca atoms with a Bi atom (8-fold coordinated) in a 2x2x1 and 2x1x2 supercells of $CaSnO_3$ and $CaZrO_3$, respectively. These supercells were used as input in FDMNES without applying any relaxation. The results for the samples and the references are shown in Figure 7c, d and Figure S3a, b. The agreement between data and simulations is very good for both samples and references, with the pre-edges and the edge region well reproduced by the calculations. The difference in the relative intensities of the features after the edge is also reproduced by the simulations, indicating that the simple structure considered catches most of the physics behind the XANES spectra. Applying relaxation would probably improve the agreement but it is beyond the scope of this work.

We can have a look to the partial DOS extracted from FDMNES to attribute the pre-edge features. The $L_3$ edge probes dipole transitions from the $2p_{3/2}$ of Bi to the unoccupied states with d and s symmetries around Bi. Comparing the simulation of $NaBiO_3$ with the projected DOS (Figure S3a) shows that the pre-edge is due to transitions to the unoccupied and sharp s-DOS close to the Fermi level. The same comparison for $Bi_2O_3$ shows that the pre-edge in this case is due to transitions to the d-DOS that in corre-spondence of the pre-edge is hybridized with the Bi p-DOS (Figure S3b). Bi in $CaZrO_3$ and $CaSnO_3$ is analogous to $Bi_2O_3$, the pre-edge reflects the hybridization between d-DOS and p-DOS. XANES analysis unambiguously confirmed the stabilization of $Bi^{3+}$ into $Ca^{2+}$ site.

## CONCLUSIONS

In this work, the luminescent properties of the Bi-activated orthorhombic perovskite $CaMO_3:Bi^{3+}$ systems ($M$=Zr, Sn, Ti) were deeply investigated. The combination of a detailed spectroscopic analysis in a wide temperature and energy scale range together with semi-empirical and theoretical models provide important insights into the origin of emitting states of the $Bi^{3+}$-doped $CaMO_3$ perovskites family. The comparison between the temperature dependence of the energy shift of the emission bands and the exciton peaks trend of the $CaMO_3:Bi^{3+}$ perovskite systems provides an unprecedent approach to shed light on the origin of the emitting states in $Bi^{3+}$-activated phosphors. HERFD-XANES spectroscopy confirmed the stabilization of $Bi^{3+}$ into $Ca^{2+}$ site and, the low temperature VUV spectroscopy, together with the design of the VRBE diagram allowed to figure out the whole picture of the energy level diagram of $Bi^{3+}$. Based on the approach reported here for the $CaMO_3$ perovskite prototype compounds, a new perspective to uncover the origin of the emitting states in $Bi^{3+}$-activated luminescent materials is demonstrated.

## METHODS

**Sample Preparation.** Bi-doped $Ca_{0.995}ZrO_3:Bi_{0.005}$ ($CaZrO_3:Bi^{3+}$), $Ca_{0.995}SnO_3:Bi_{0.005}$ ($CaSnO_3:Bi^{3+}$) and $Ca_{0.995}TiO_3:Bi_{0.005}$ ($CaTiO_3:Bi^{3+}$) samples were prepared by conventional solid-state reaction method. For the charge neutrality, cation vacancies can be generated. The chemical reagents $CaCO_3$ (4N), $ZrO_2$ (4N), $SnO_2$ (4N), $TiO_2$ (4N) and $Bi_2O_3$ (4N) were used as starting materials, grounded in an alumina mortar to form homogeneous fine powder mixtures of the desired composition. The mixtures were calcined at 800 °C for 3 h, cooled, grounded again and pressed into pellets (0.45 g, $\phi$ 15 mm) with a uniaxial loading of 50 MPa in a stainless steel mold. Then, the pellets were fired at 1300 °C, 1200 °C and 1100 °C for 6 h in air atmosphere for the $CaZrO_3:Bi^{3+}$, $CaSnO_3:Bi^{3+}$ and $CaTiO_3:Bi^{3+}$, respectively.

**Experimental Details.** The crystal phase was identified by XRPD measurement (Shimadzu, Kyoto, Japan; XRD6000). The diffuse reflectance spectra were collected by a spectrophotometer (Shimadzu, UV3600) equipped with an integrating sphere.

**Optical Spectroscopy.** PLE spectra were collected exciting with a 300 W Xe lamp (Asahi Spectra, MAX-302) equipped with a monochromator (Nikon, G250) and detecting by Si photodiode (PD) detector (Electro-Optical System Inc., S-025-H) equipped with a monochromator (Shimadzu, 675 grooves/mm). PL spectra were measured collecting with a CCD spectrometer (Ocean Optics, QE65Pro) connected with an optical fiber. Temperature dependence of PL and PLE (80-600 K) were investigated setting the sample in a cryostat (Helitran LT3, Advanced Research Systems). The temperature dependence PL spectra in the 20-300 K range were collected by means of a closed circuit He cryostat (Iwatani industrial gases Co., Ministat). All the spectra were calibrated by means of a standard halogen lamp (Labsphere, SCL-600).

**Synchrotron-radiation VUV Spectroscopy.** PLE spectra in the vacuum ultraviolet (VUV) and visible region (100-500 nm) were measured at the BL3B beamline of the UVSOR facility (Institute for Molecular Science, Okazaki, Japan) under helium cooling (9K). The beamline consists of a 2.5 m off-plane Eagle type normal incidence monochromator, which covers the VUV, UV and visible regions. In the present experiments a spherical grating with a groove density of 300 lines/mm optimized at a photon energy of ~12 eV was used.[78] High-order light from the normal incidence monochromator was removed using lithium fluoride and quartz plates, and colored glass filters. PLE spectra obtained were corrected for the spectral distribution of the excitation light source. It is important to point out that for a reliable calculation of the peak position, the spectra were considered in energy scale after suitable transformation.

**XANES Spectroscopy.** Bi $L_3$-edge HERFD-XANES spectra were recorded at beamline ID26 at the European Synchrotron Radiation Facility. The incoming energy was selected by a Si(311) double crystal monochromator, the ID26 emission spectrometer was equipped with a set of 5 Ge(844) bent crystals of 1 m radius to select the maximum of the Bi $L\alpha_1$ florescence line. The X-ray spot size at the sample position was 600 x 80 $\mu$m (horizontal x vertical). XANES analysis was conducted comparing the experimental spectra to calculations obtained with the FDMNES code. The atomic coordinates for Bi in $CaZrO_3$ and $CaSnO_3$ were obtained by substituting a Ca atom with a Bi atom in a supercell of 2x2x1 of $CaSnO_3$ and a supercell of 2x1x2 of $CaZrO_3$. Relativistic and spin-orbit effects as well as quadrupolar transitions were considered. The Finite Difference Method (FDM) was used in a cluster of 6 A around the absorber (70 atoms). The parameters for convolution were adapted to have the correct broadening of post-edge features.

## ASSOCIATED CONTENT

**Supporting Information**. Normalized temperature dependent PL spectra, synchrotron-radiation VUV-Vis PLE spectra deconvolution analysis (9 K) and experimental Bi $L_3$-edge HERFD-XANES together with the results of FDMNES calculations and the partial DOSs extracted for Bi in the $Bi_2O_3$ and $NaBiO_3$ references. This material is available free of charge via the Internet at http://pubs.acs.org.

## AUTHOR INFORMATION

### Corresponding Authors

* E-mail: michele.back.68r@st.kyoto-u.ac.jp (M.B.)
* E-mail: tanabe.setsuhisa.4v@kyoto-u.ac.jp (S.T.)

### Author Contributions

The manuscript was written through contributions of all authors. All authors have given approval to the final version of the manuscript.

### Funding Sources

Any funds used to support the research of the manuscript should be placed here (per journal style).

**Notes**

The authors declare no competing financial interest.

## ACKNOWLEDGMENT

Work by M. B. was financially supported by the Grant-In-Aid for JSPS Fellows (17F17761). We thank UVSOR staff members for their technical support. This work was supported by the UVSOR Facility Program (Project number: 30-822, BL-3B) of the Institute for Molecular Science and the Grant-in-Aid for Scientific Research on Innovative Areas "Mixed Anion" project (16H06441) from MEXT. L. A. acknowledges the ERC under grant agreement No 759696 for financial support. The authors would like to thank the ID26 team and T. Bohdan for technical support during beamtime and Jonas J. Joos for providing Bi references.

**TOC Graphic**

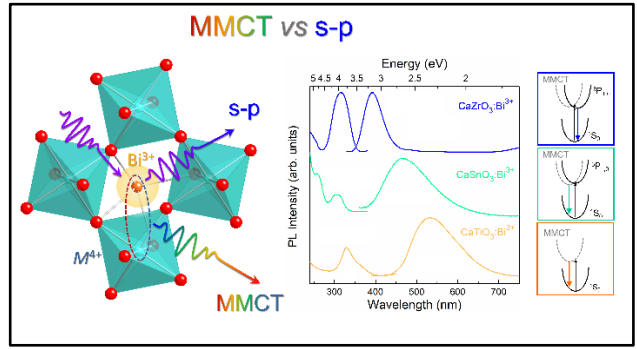